# Electronic Structure of Ordered Double Perovskite $Ba_2CoWO_6$

Rajyavardhan Ray[1,2], A K Himanshu*[2], Kumar Brajesh[3], B K Choudhary[4], S K Bandyopadhyay[2], Pintu Sen[2], Uday Kumar[5], and T P Sinha[6]

[1]*Department of Physics, Indian Institute of Technology Kanpur, Kanpur-208016*
[2]*Nanostructured & Advanced Material Laboratory, Variable Energy Cyclotron Centre, 1/AF, Bidhannagar, Saltlake, Kolkata-700064*
[3]*Department of Physics, Veer Kunwar Singh University, Ara, Bihar, India*
[4] *University Department of Physics, Ranchi University, Jharkhand, India*
[5] *Department of Physical Sciences, IISER, Mohanpur Campus, Mohanpur, West Bengal, India*
[6] *Department of Physics, Bose Institute, Kolkata, India*

*himanshu_ak@yahoo.co.in, akh@vecc.gov.in (Email of corresponding author)

**Abstract.** $Ba_2CoWO_6$ (BCoW) has been synthesized in polycrystalline form by solid state reaction at 1200 ˚C. Structural characterization of the compound was done through X-ray diffraction (XRD) followed by Rietveld analysis of the XRD pattern. The crystal structure is cubic, space group Fm-3m (No. 225) with the lattice parameter, a = 8.210. Optical band-gap of the present system has been calculated using the UV-Vis Spectroscopy and Kubelka-Munk function, it's value being 2.45 eV. A detailed study of the electronic properties has also been carried out using the density functional theory (DFT) techniques implemented on WIEN2k. Importance of electron-electron interaction between the Co ions leading to half-metallic behavior, crystal and exchange splitting together with the hybridization between O and Co, W has been investigated using the total and partial density of states.

Keywords: **Double Perovskite, Optical band-gap, Kubelka-Munk function, Electronic Structure, Density Functional theory.**

**PACS:** 31.15A, 31.15.E, 75.50Ee

## INTRODUCTION

Double Perovskite, a broad class of materials with the general formula $A_2BB'O_6$, have attracted a lot of attention since the 1960s [1,2] as they exhibit a variety of exotic properties like high temperature superconductivity, collosal magnetoresistance, half-metallicity, and magnetodielectricity. In particular, the half-metallicity, characterized by differentiated conducting response of the spin up and spin down orientations, has attracted a lot of attention due to possible applications in spintronic devices. A-site ion can be an alkaline earth metal such as Barium, Strontium, Calcium or a Lanthanide, and B and B' are usually transition metals. Each B-site cation (B and B') is surrounded by an Oxygen octahedron, and the A-site atoms are situated in the corner produced by the eight adjacent Oxygen octahedral. The origin of half-metallicity is typically attributed to the magnetic and non-magnetic cations at the B and B' site respectively.

Cobalt-tungstates, with the general formula $A_2CoWO_6$ (A = Sr, Ba, Ca) are among the largely studied half-metallic double perovskites. While there is much literature available for the Sr and Ca cobalt tungstates, there is no study of electronic structure of the Ba counterpart. Barium Cobalt Tungstate (BCoW) is known to be antiferromagnetic with $T_N$=19K. The crystal structure is cubic Fm-3m with lattice constant a=8.098Å [3]. In this article, we, therefore, report the optical band-gap of Barium cobalt tungstate obtained through the UV-Vis reflectance spectroscopy and carry out an *ab-initio* study of the electronic properties.

## EXPERIMENTAL AND CALCULATION DETAILS

Samples of BCoW were synthesized using the solid-state reactions of stoichiometric mixtures of BaCO3, CoO and $WO_3$ heated at 1200 °C for eight hours. Color of the obtained sample was found to be yellowish dark green. X-ray diffraction was carried out and was analyzed using Rietveld analysis software FULLPROF. The structure is found to be cubic, space group Fm-3m (No. 225) with lattice constant a=8.210Å. The positions of Ba, Co, W, and O atoms are found to be *8c*, *4a*, *4b* and *24e* respectively with the x-coordinate for the O atom in *24e* position equal to 0.266.

In order to study the optical band-gap of the sample, diffusive reflectance measurement was carried out through UV-Vis spectroscopy (Perkin-Elmer 950). The Kubelka-Munk function was used to convert the reflectance spectra, thus obtained, into an equivalent absorption spectra [4,5]. The optical band-gap was obtained from the linear part of the energy dependence of the absorption spectra using an appropriate function, corresponding to direct-allowed transition. FIGURE 1 shows the absorption spectra and the corresponding value of band-gap. The value of the optical band gap was found to be approximately 2.46eV, which is in accordance with the color of the synthesized sample, which clearly suggests a band-gap > 2.4eV.

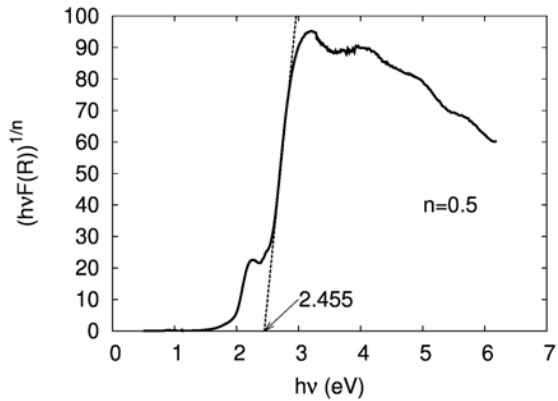

**FIGURE 1.** Linear dependence of the modified Kubelka-Munk function F(R) on the photon energy hυ corresponding to direct-allowed transitions, showing and optical band-gap of 2.46eV.

Calculation of the band and electronic structure for this complex double perovskite were performed using the Full Potential - Linear Augmented Plane Wave (FP-LAPW) method, in the framework of first principles Density Functional Theory (DFT), as implemented in WIEN2k. Starting with the experimentally obtained structure and atomic positions, volume optimization has been performed to obtain the fully-relaxed structure, obtained from the Birch-Murnaghan equation of state [5]. FIG. 2 shows the values of the energy corresponding to different values and well as the Birch-Murnaghan equation of state (dashed line). The values of optimal volume is found to be 934.06 a.u.$^3$. We use 72 k-points in the Brillouin zone and the muffin-tin radii for Ba, Co, W, and O are, respectively, obtained to be 2.5, 2.22, 1.96 and 1.73. The density plane cut-off $R*k_{max}$ is 7.0, where $k_{max}$ is the plane-wave cut-off and R is the smallest of all atomic radii. The exchange and correlation effects have been treated within the Generalized Gradient Approximation (GGA). The self-consistency is better than 0.001 e/a.u.$^3$ for charge and spin density, and the stability is better than 0.01 mRy for the total energy per unit cell.

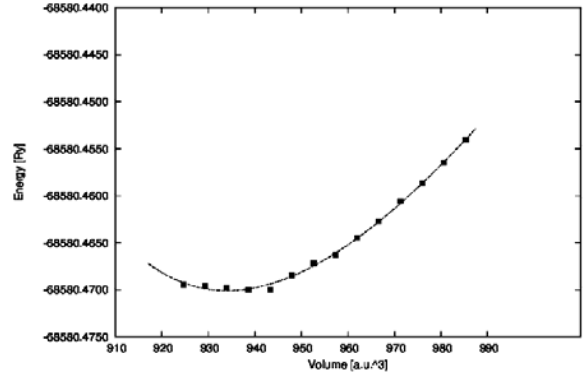

**FIGURE 2.** Energy versus volume for $Ba_2CoWO_6$. The dashed curve is the Birch-Murnaghan equation of state.

## RESULTS AND DISCUSSION

The spin-polarized total density of states (DOS) per unit cell of BCoW reveals the system to be metallic which is contrary to the experimental observation above. However, with even a relatively small value of the electron-electron interaction (Coulomb repulsion term U), the system exhibits half-metallic behavior with an insulating gap in the spin-up channel and a conductor feature for the spin down channel. The DFT calculations were performed for a range of U-values between 0.15eV and 3eV. The appropriate values of U which correspond to the experimentally determined band gap is found to be U= 0.5eV, implying weak correlations for the electrons in this system.

FIGURE 3 shows the total and partial density of states for U=0.5eV, clearly exhibiting the half-metallic nature with a spin gap of 2.5eV through the Fermi level for spin up channel. There are states of Co in both spin up and down channels of the valence band and is responsible for this half-metallic effect in this double perovskite compound. On the other hand, the W ion contributes to the conduction band. States close to the Fermi energy in the conduction band, however, has dominant contribution from the W ion in the spin up channel and Co ion in the spin down channel. While in the spin up channel, there is strong hybridization between O and Co ions in the valence band and relatively weak bonding between O and W in the conduction band, from evidences spin down channel show bonding between O and W ions in the conduction band.

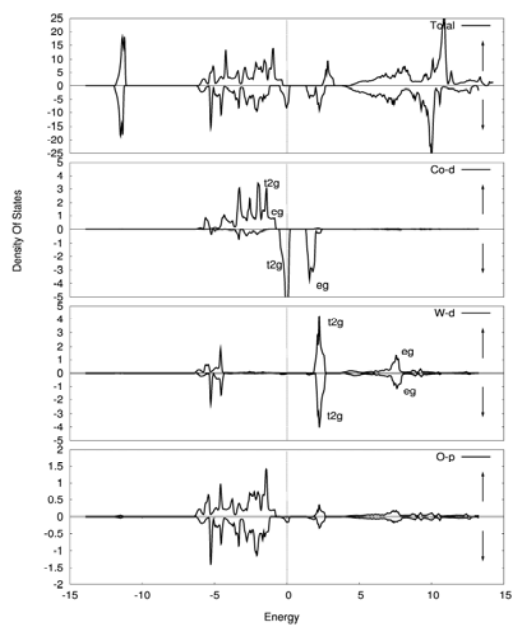

**FIGURE 3.** Total and partial spin-polarized density of states of Co-d, W-d and O-p in $Ba_2CoWO_6$.

The crystalline field produced by the oxygen octahedra lifts the degeneracy and leads to splitting of the d-states of Co and W atoms into $t_{2g}$ and $e_g$ states. Exchange splitting between the $t_{2g}$ up and $t_{2g}$ down, and between $e_g$ up and $e_g$ down states are also observed for the Co ion. For the W ion, on the other hand, there is no exchange splitting due to complete overlap of the spin up and spin down states. This indicates the non-magnetic behavior of W, as also suggested by the 6+ valency ($5d^0$ occupancy) in this compound. The exchange splitting between the $e_g$ up and $e_g$ down states and between $t_{2g}$ up and $t_{2g}$ down states for Co is 2.6eV and 1.4eV respectively, and the crystal field splitting between the $t_{2g}$ and $e_g$ Co states in spin channel up and down is 0.48eV and 1.73eV respectively. In the spin up channel, both $t_{2g}$ and $e_g$ states of Co are fully occupied in comparison to the spin down channel, where the $t_{2g}$ states of Co are partially filled and $e_g$ states are completely empty. For W, both $t_{2g}$ and $e_g$ are completely empty as expected.

From the above analysis, it is also found that the occupation for Co ion is $t_{2g}^5 e_g^2$ which indicates the 2+ valency of Co ions and, therefore, contributes to the magnetic moments in this compound. The magnetic moment per unit cell contribution from Co is found to be 2.62 $\mu_B$ which is in close agreement with the previously obtained values [7] The total magnetic moment per unit cell is found to be 3.004 $\mu_B$ as expected for $Co^{2+}$. This is, however, significantly smaller than the previously reported values [3,8].

## CONCLUSIONS

In conclusion, we have experimentally obtained the optical band-gap and theoretically studied the electronic structure of BCoW using FP-LAPW based first principles method. The ground state is half-metallic with the spin up channel being insulating with a spin gap of 2.5eV. Although the electron-electron interaction between the Co ions is small, it is significant and necessary for the half-metallic behavior in the present compound.